%% file: Island_Interactions.tex
\documentclass[prl, twocolumn, superscriptaddress]{revtex4-1}
\usepackage{graphicx}

\usepackage{float}
\usepackage{amsmath}
\usepackage{ragged2e}

\begin{document}
\title{Interaction-driven giant orbital magnetic moments in carbon nanotubes}
\author{Joshua O. Island}
\email{jisland@physics.ucsb.edu}
\altaffiliation[Present address: ]{Department of Physics, University of California, Santa Barbara CA 93106 USA.}
\affiliation{Kavli Institute of Nanoscience, Delft University of Technology, Lorentzweg 1, 2628 CJ Delft, The Netherlands.}
\author{Marvin Ostermann}
\affiliation{Kavli Institute of Nanoscience, Delft University of Technology, Lorentzweg 1, 2628 CJ Delft, The Netherlands.}
\author{Lee Aspitarte}
\affiliation{Department of Physics, Oregon State University, Corvallis, Oregon 97331, United States.}
\author{Ethan D. Minot}
\affiliation{Department of Physics, Oregon State University, Corvallis, Oregon 97331, United States.}
\author{Daniele Varsano}
\affiliation{CNR-NANO, Via Campi 213a, 41125 Modena, Italy.}
\author{Elisa Molinari}
\affiliation{CNR-NANO, Via Campi 213a, 41125 Modena, Italy.}
\affiliation{Dipartimento di Scienze Fisiche, Informatiche e Matematiche (FIM), Universit\`a degli Studi di Modena e Reggio Emilia, 41125 Modena, Italy.}
\author{Massimo Rontani}
\email{massimo.rontani@nano.cnr.it}
\affiliation{CNR-NANO, Via Campi 213a, 41125 Modena, Italy.}
\author{Gary A. Steele}
\affiliation{Kavli Institute of Nanoscience, Delft University of Technology, Lorentzweg 1, 2628 CJ Delft, The Netherlands.}

\begin{abstract}
Carbon nanotubes continue to be model systems for studies of confinement and interactions. This is particularly true in the case of so-called ``ultra-clean'' carbon nanotube devices offering the study of quantum dots with extremely low disorder. The quality of such systems, however, has increasingly revealed glaring discrepancies between experiment and theory. Here we address the outstanding anomaly of exceptionally large orbital magnetic moments in carbon nanotube quantum dots. We perform low temperature magneto-transport measurements of the orbital magnetic moment and find it is up to seven times larger than expected from the conventional semiclassical model. Moreover, the magnitude of the magnetic moment monotonically drops with the addition of each electron to the quantum dot directly contradicting the widely accepted shell filling picture of single-particle levels. We carry out quasiparticle calculations, both from first principles and within the effective-mass approximation, and find the giant magnetic moments can only be captured by considering a self-energy correction to the electronic band structure due to electron-electron interactions. 
\end{abstract}
\maketitle

A steady increase in the quality of carbon nanotube (CNT) devices has lead to a deeper understanding of the physics that governs this material system that has captivated researchers for over two decades. This is particularly exemplified in the 2005 work by Cao \textit{et al.} when they presented a method to fabricate ultra-clean carbon nanotube transport devices whereby the nanotube was grown in the last step of fabrication\cite{cao2005electron}. This method greatly alleviated disorder brought on by defects, absorbed contaminants, and the underlying substrate\cite{amer2012influence, charlier2002defects}. The quality of similarly fabricated devices has lead to observations of elegant subtleties beyond early measurements of single electron tunneling such as an intimate coupling between spin and orbital motion\cite{kuemmeth2008coupling}, Wigner crystallization\cite{deshpande2008one, pecker2013observation}, and strong feedback between electron tunneling and mechanical motion\cite{steele2009strong}. While these experiments are a testament to the quality of ultra-clean devices, they have increasingly offered glimpses of anomalous behavior which seem to persist without explanation\cite{laird2015quantum}. 

In 2004, the orbital magnetic moment of electrons circling a carbon nanotube was shown to be a simple function of the nanotube diameter ($D$) and the electron's Fermi velocity ($v_F$),  $\mu_{orb} = Dev_F/4$ (Ref. \onlinecite{minot2004determination}). This relation has been supported by other works finding reasonable agreement between magneto-transport measurements of $\mu_{orb}$ and measurements of the average nanotube diameter for certain growth conditions\cite{makarovski20072, churchill2009electron}. Some works however find deviations from this relation where either measured $\mu_{orb}$'s infer exceptionally large single wall nanotube diameters\cite{jarillo2005electronic, kuemmeth2008coupling, jespersen2011gate} ($> 3$ nm) or the measurements of $\mu_{orb}$ and the nanotube diameter simply do not agree at all\cite{steele2013large}. Deviations in the Fermi velocity, with reported experimental values of $(0.8 - 1.1) \times 10^6$ ms$^{-1}$ (Refs. \onlinecite{lemay2001two, chuang2008cyclotron}), cannot account for the disagreement. 

Reports of the magnitude of spin orbit coupling in carbon nanotube quantum dots have shown a similar trend. Spin orbit couplings as large as sixteen times greater than theoretical predictions have been measured\cite{jhang2010spin, steele2013large}. A fraction of this discrepancy may lie in the use of the measured orbital magnetic moment to determine the nanotube diameter which, as we note above, can lead to discrepancies. Theory predicts Zeeman-like and orbital-like spin-orbit couplings of $\delta_{SO}^0\approx - \frac{0.3\text{ meV}}{D\text{ (nm)}}cos\text{ }3\theta$ and $\delta_{SO}^1\approx - \frac{0.3\text{ meV}}{D\text{ (nm)}}$, respectively, $\theta$ being the chiral angle\cite{laird2015quantum, maslyuk2017spin}. A larger inferred nanotube diameter would invariably lead to smaller theoretically predicted spin-orbit couplings. 

Lastly, one of the longstanding mysteries in low temperature transport experiments on carbon nanotubes is a non-closing or residual band gap at the Dirac field (closing field) for quasi-metallic (small band gap) nanotubes. Theory says that metallic nanotubes can develop a band gap due to symmetry breaking of the underlying graphene lattice from strains, twists and curvature. The magnitude of this gap is predicted to be around tens of milli-electron volts but zero field gaps of an order of a magnitude larger have been reported\cite{island2011ultra, aspitarte2017giant, mcrae2017giant}. Perhaps most intriguingly though, in the single particle picture, these gaps should vanish at the Dirac field as the nanotube quantization line is pushed to the Dirac point of the underlying graphene band structure resulting in a truly metallic nanotube. In experiment this has not been observed and typically a residual gap exists of tens of milli-electron volts. Deshpande et al. have interpreted this phenomena in the context of a Mott insulating phase\cite{deshpande2009mott}. The extracted $1/R^{1.3}$ dependence, where $R$ is the nanotube radius, however relied on inferred nanotube diameters from the measured orbital magnetic moments. 
\clearpage

\onecolumngrid

\begin{figure}[h]
\includegraphics [width=6 in]{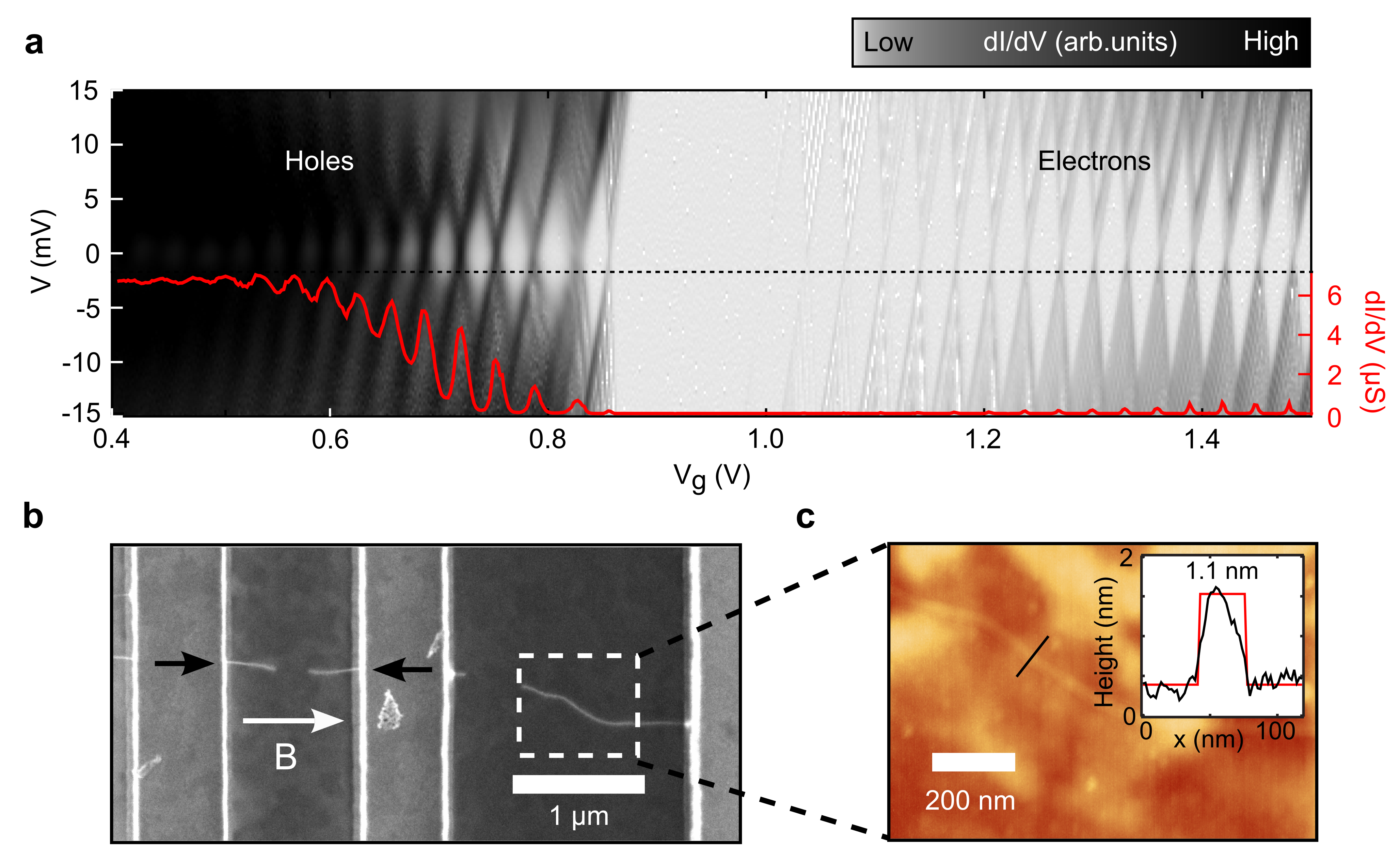}
\caption{\label{} (a) Grey-scale plot of (d$I$/d$V$) as a function of bias voltage ($V$) and gate voltage ($V_g$) at 3 K. The overlaid line cut shows (d$I$/d$V$) at a bias voltage of $V \approx -2$ mV. (b) Scanning electron microscopy (SEM) image of device A. Note that this tube ruptured after measurements and before imaging. Two trenches can be seen, one on the left where the nanotube is indicated with two black arrows and one on the right which was used for height analysis. (c) Atomic force microscopy (AFM) image of the same tube. The inset shows a linecut across the tube at the location of the black line in the AFM image. }
\end{figure}

\twocolumngrid

Recent experiments from the present authors and others have shown that the nanotube band gap is extremely sensitive to the dielectric environment supporting indications of a strong role from interactions\cite{lin2010many, aspitarte2017giant}. Emerging theory suggests the residual gap in narrow-gap nanotubes is the manifestation of an excitonic insulating phase, stabilized in the ground state by long-range Coulomb interaction, as electron-hole pairs spontaneously condense near the Dirac field where the transport gap should completely close\cite{rontani2014anomalous, Varsano2017}. The experimental signature of this exciting phase predicted over 50 years ago\cite{mott1961transition, knox1963solid, keldysh1965possible, jerome1967excitonic, sherrington1968speculations} is a slower decay of the residual gap as a function of the nanotube radius ($\Delta_{res}\sim1/R$) as compared with the predicted Mott gap decay ($\Delta_{res}\sim1/R^{1/(1-g)}$, with $g<1$)(Ref. \onlinecite{Varsano2017}). The true nanotube radius would be required to differentiate the two paradigms and elucidate the origin of the non-closing gap.

Here, we report on orbital magnetic moments in ultra clean carbon nanotube quantum dots that deviate from existing theory both qualitatively and quantitatively. Instead of a magnetic moment which remains constant within a shell, we find that the orbital magnetic moment decreases monotonically with each added electron. Additionally, we analyze the magnitude of the moments and find that they are much larger than expected from semiclassical estimates based on a direct measurement of the nanotube diameter. We further compare our results with other models taking into account a change in the size of the quantum dot with filling, a change in charging energy with magnetic field, and the orbital magnetic moment of a Wigner molecule. None of these models suffice to explain the magnitude or trend of our observations. It is only by treating electrons added to the dot as quasiparticles dressed by the Coulomb interaction with other electrons already present in the nanotube, including those in the filled valence band, that we are able to account for the enhanced orbital magnetic moment. A self-energy correction to the gap computed within the effective-mass approximation results in good agreement between observations and theory and is further validated by a first-principles GW calculation for a small nanotube. Finally, in agreement with previous studies, we show that our small band gap tubes present a residual gap at the Dirac field and we discuss the implications of our results in the context of the Mott and excitonic insulating phases. 

The fabrication of our ultra-clean suspended devices follows from the process developed in Ref. \onlinecite{schneider2012coupling} using a methane growth recipe detailed in Ref. \onlinecite{steele2009tunable}. Figure 1(b) shows a scanning electron microscopy image of a characteristic device (device A) after measurement. Note that this tube ruptured after measurement and before imaging. The nanotube can be seen at the location of the black arrows for the left trench. Low temperature measurements are performed in a dilution fridge at temperatures of 100 mK - 4 K. We measure the current ($I$) as a function of two terminal voltage bias ($V$), back-gate voltage ($V_g$), and magnetic field ($B$). We measured four devices (labeled A-D) in detail having similar low temperature characteristics (see the Supplemental materials for additional devices\cite{SUPP}\nocite{Ando1997, Giannozzi2009, Perdew1981, Miyake2005, Marini2009, Godby1989, Bruneval2008, Rozzi2006, Sangalli2011, Abramowitz1972, Giuliani2005, Spataru2012}) and present one device (A) in the main text for consistency.

We now turn to the low temperature measurements of our carbon nanotube devices. Figure 1(a) shows a stability diagram of the calculated differential conductance (d$I$/d$V$) (from the measured current ($I$)) as a function of bias voltage, $V$, and back-gate voltage, $V_g$, for device A. The characteristic Coulomb blockade diamonds can be observed signaling the single electron transistor (SET) regime with a stable confinement of holes (on the left) and electrons (on the right) separated by a small energy band gap. The well-defined periodicity of the Coulomb diamonds and uniformity of the slopes indicate transport through a single defect-free carbon nanotube quantum dot. We observe orders of magnitude larger currents for holes than electrons (overlaid line profile in Figure 1(a)) due to hole doping from the electrodes\cite{schneider2012coupling}. An estimate of the band gap can be made from subtracting the average addition energies (heights of diamonds on the left and right of the band gap, see Supplemental Materials for a stability plot of low filling\cite{SUPP}) for the first hole and electron from the height of the central diamond\cite{kuemmeth2008coupling}. For device A we estimate a zero field gap of $\approx 76$ meV. 

\begin{figure}
\includegraphics [width=3 in]{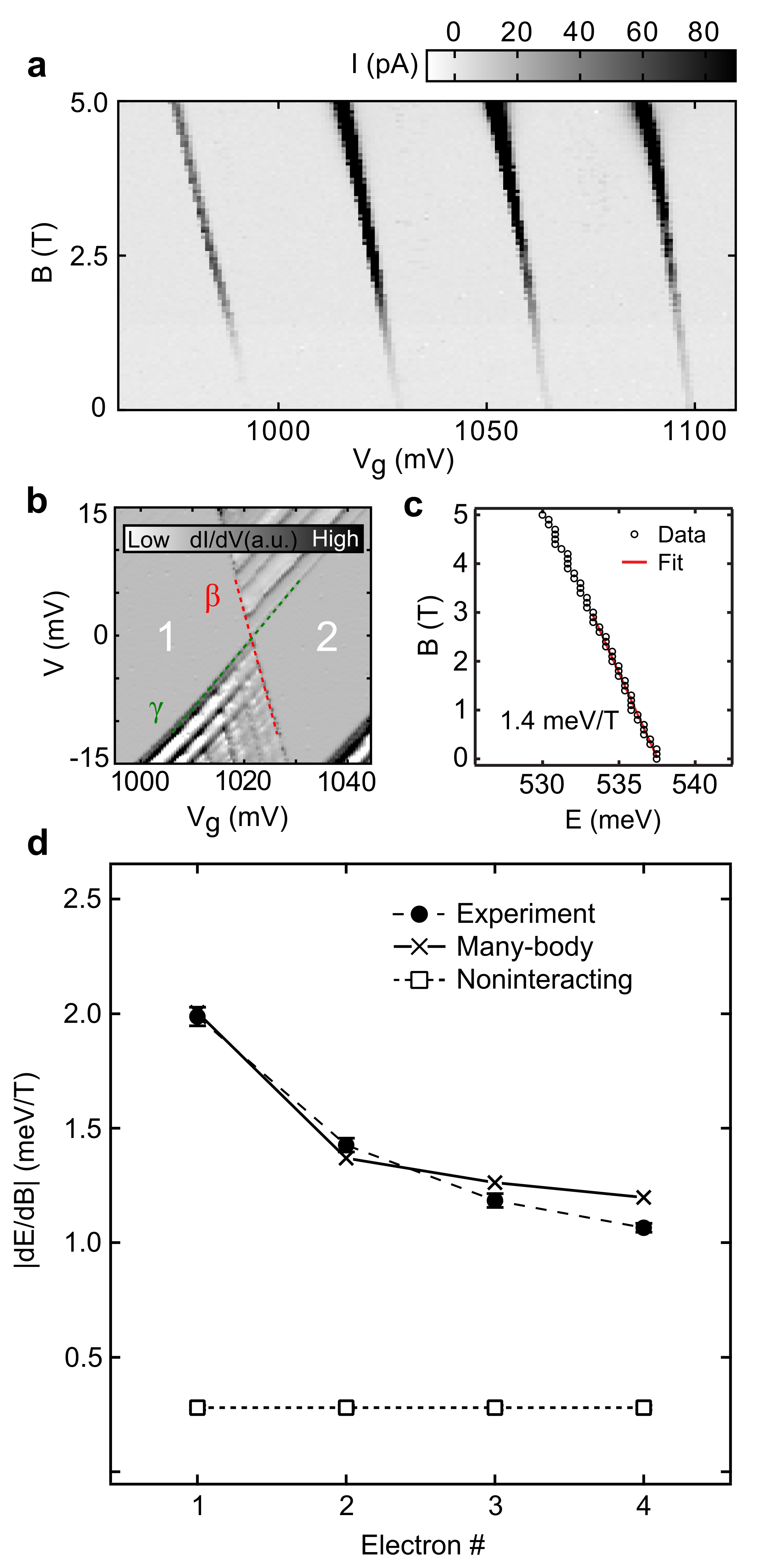}
\caption{\label{} (a) Measured current ($I$) at a bias voltage of 1 mV as a function of magnetic field. The first four Coulomb peaks are shown corresponding to the ground state for the first four electrons (the first shell) for device A (T = 100 mK). (b) Extraction of the voltage-to-energy conversion ($\alpha$-factor) from the Coulomb diamonds measured at 100 mK. (c) $\alpha$-factor converted energy of the second electron as a function of field. The orbital magnetic moment is extracted by taking the slope of the change in the ground state energy with magnetic field. (d) Orbital magnetic moment for the first four electrons for device A. Black circles show the magnitude of the orbital magnetic moments extracted from the data in (a). Open squares show the semiclassical estimates from the measured diameter of the nanotube for device A (shown in Fig. 1(c)). Crosses mark the results from our effective-mass GW calculations.}
\end{figure}

In the simplest picture, the electronic states in carbon nanotube quantum dots can be thought of as semiclassical orbits around the circumference of the tube giving circling electrons on the nanotube an orbital magnetic moment of $\mu_{orb}=Dev_F/4$ directed along the nanotube axis\cite{minot2004determination}. Upon application of a magnetic field along the tube axis, this orbital magnetic moment causes a shift in the energy of the electronic states of $\Delta E=-\boldsymbol{\mu}_{orb}\cdot \textbf{B}=\pm \mu_{orb}B_{\parallel}$. The shift is either negative or positive depending on the orientation, clockwise or anticlockwise, of the circular orbits which correspond to electrons in the \textbf{K} or \textbf{K'} valley of the electronic band structure\cite{minot2004determination}. In the single particle shell filling picture, two electrons (spin up and spin down) fill the lowest energy valley and the subsequent two electrons fill the next valley giving a total of four electrons per shell\cite{liang2002shell, sapmaz2006quantum}. A straightforward estimate of the magnitude of the expected orbital magnetic moment for a carbon nanotube quantum dot can be made by measuring directly the nanotube diameter. For example, in Fig. 1(c) we show an atomic force microscopy (AFM) image of device A taken at the location indicated by the box in the scanning electron microscope (SEM) image in Fig. 1(b). For this tube having a relatively small diameter of $1.10\pm0.03$ nm (from five independent measurements of the tube), we expect an orbital magnetic moment of $\mu_{orb} = 0.28\pm0.01$ meV/T for the first shell (see the Supplemental Material\cite{SUPP} for another device (B) having a larger diameter of $3.00\pm0.04$ nm and expected $\mu_{orb}$ of $0.75\pm0.01$ meV/T). 

In order to directly extract the experimental magnitude of the orbital magnetic moment, we apply a parallel component of the magnetic field along the nanotube axis (indicated in Fig. 1(b)) and measure the change in the energy of the ground state. The first four Coulomb peaks corresponding to the first shell of device A are shown in Fig. 2(a) and plotted as a function of magnetic field. The shift in energy of each level is related to the gate voltage through the factor $\alpha=|e|C_g/C_{tot}$. Fig. 2(b) shows the single electron tunneling (SET) regions for the first and second electronic ground states. The gate coupling is related to the positive ($\gamma$) and negative ($\beta$) slopes of the SET regions by, $1/\alpha = 1/\beta + 1/\gamma$, with $\beta=|e|C_g/C_s$ and $\gamma=|e|C_g/(C_{tot}-C_s)$, and where $C_g$, $C_s$, and $C_{tot}$ are the gate, source, and total capacitances of the system. For the ground state charged with two electrons we calculate a gate coupling of $\alpha=0.52$ eV/V. Using this coupling we plot the peak position, in energy, for the second electron from the data in Fig. 2(a) in Fig. 2(c). From a linear fit of this data from 0 to 3 T we estimate an orbital magnetic moment of $\mu_{orb}\approx|$d$E$/d$B|=1.42\pm0.03$ meV/T. The error here is to account for a possible deviation of $\pm10^\circ$ in the parallel component of the magnetic field. 

In Fig. 2(d) we plot the measured $|$d$E/$d$B|$ (black filled circles) for the rest of the first shell of electrons. A maximum of $2.00\pm0.04$ meV/T is reached for the first electron and a monotonic decrease for subsequent filling of the first shell is observed. Not only is the absence of a switch to positive magnetic moment noted in Fig. 2(a) (i.e. the electrons seem to fall into one single valley), the magnitudes are much larger than expected. From the measured diameter, we estimated an orbital magnetic moment of $\mu_{orb} = 0.28$ meV. This is seven times smaller than the measured d$E/$d$B$ for the first electron. We note that the expected Zeeman contribution of $\pm(1/2)g\mu_BB_\parallel=\pm0.058B_\parallel$ (meV) does not make up for the difference. In Fig. 2(d) (open squares) we plot the magnitude in energy for the next three electrons as well which are expected to stay constant within the shell. There is a clear disagreement between the single particle model and the measured orbital magnetic moment. The measured moment for the first electron would correspond to a nanotube with a diameter of 8 nm in the semiclassical picture which exceeds the theoretical collapse threshold for single walled nanotubes of 5.1 nm\cite{he2014precise}. In addition, chemical vapor deposition grown nanotubes rarely exceed 3 nm in diameter\cite{kong1998chemical, chen2014diameter}. The disagreement between the data and semiclassical estimates from the measured nanotube diameter is quite remarkable and encourages further investigation. In the Supplemental Materials we first try to recover the enhancement through modifications of the semiclassical model given changes in the size of the quantum dot or the charging energies with magnetic field\cite{SUPP}. Neither effects account for our observed enhancement. The appearance of all four electrons filling a single valley is an indication of strong electron-electron interactions and the possible formation of a Wigner molecule\cite{deshpande2008one, secchi2010wigner, pecker2013observation}. We consider a simple model of electrons in the Wigner crystal regime in the supplement which again fails to reproduce our results\cite{SUPP}. We additionally note that three of the four devices show Wigner-like characteristics indicating strong interactions and one (device B) displays single-particle-like filling but still presents an enhanced orbital magnetic moment underlying the ubiquity of our results and failure of these simple models to reproduce them.  

\begin{figure}
\includegraphics [width=3 in]{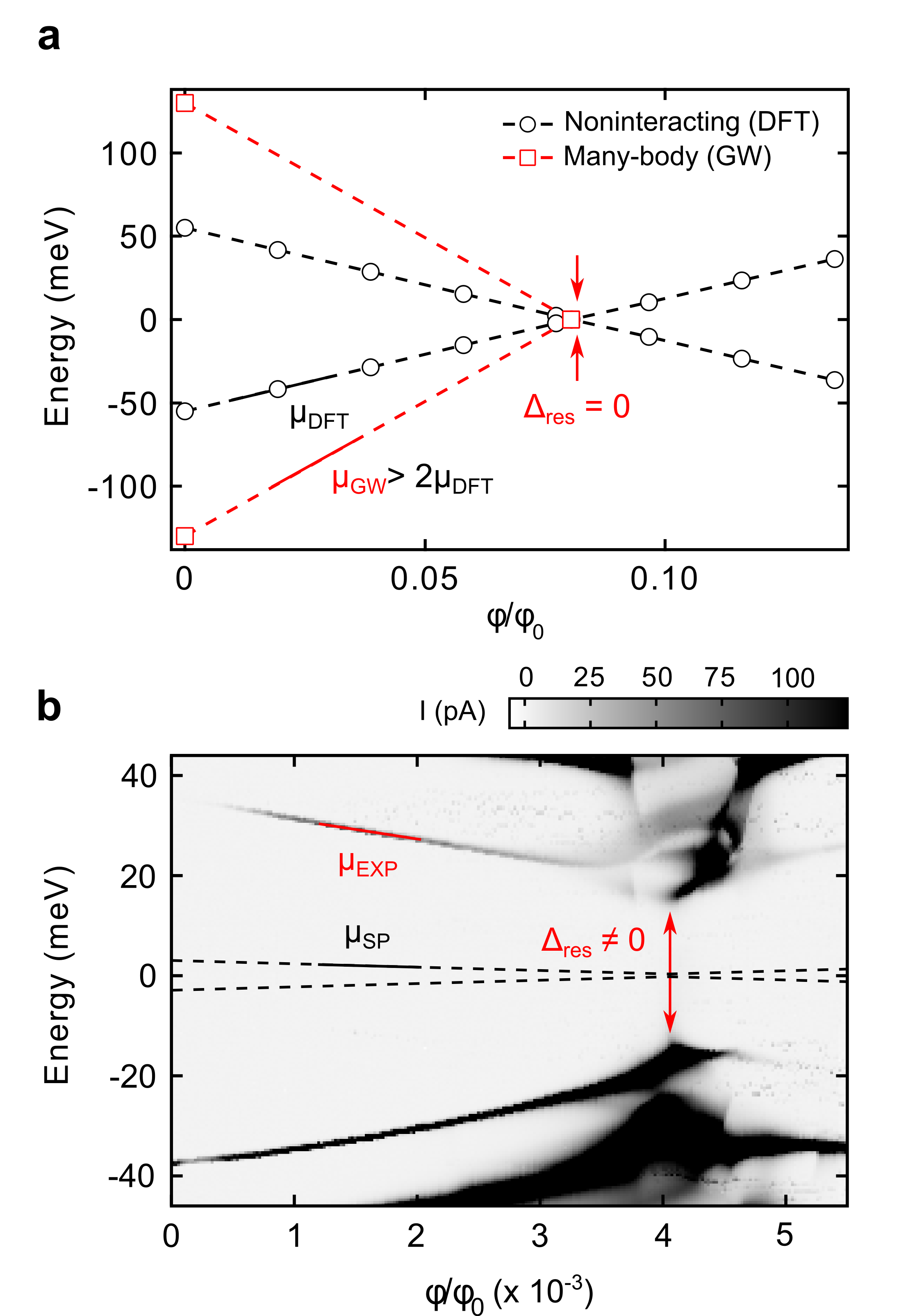}
\caption{\label{}(a) The ground state energy for the first electron (positive energy) and hole (negative energy) as a function of magnetic flux threading the nanotube to the flux quantum ($\phi_0$) calculated from first principles for the narrow-gap (9,0) zigzag tube at the density functional theory (black open circles, 'noninteracting') and GW (red open squares, 'many-body') level. (b) Measured current for device A as a function of energy (converted from gate voltage using the $\alpha$-factor) and magnetic flux for the first electron (positive energy) and hole (negative energy).}
\end{figure}

Instead, we adopt a different approach and consider the many-body correction to the noninteracting band gap induced by the quasiparticle self-energy that originates from electron-electron interactions\cite{Hedin1969, Onida2002}. We calculate the magnitude of the orbital magnetic moment within the GW scheme\cite{Hedin1969, Onida2002} and validate our predictions based on the effective mass approximation by investigating an additional case study from first principles (see Fig. 3(a) and Supplemental Materials\cite{SUPP}). We remark that our theory fully takes into account the gap-opening effect of tube curvature. The results of our effective mass calculations are shown in Fig. 2(d) along with the measured data for device A. Good agreement is found for both the magnitude and trend of the measured data. As electrons are added to the empty conduction band, the Coulomb interaction is effectively screened by the metal-like 1D Lindhard dielectric function leading to a monotonic decrease in the magnitude of the orbital magnetic moment (see Supplemental Materials for details\cite{SUPP}). The qualitative phenomenon that we have observed, the dramatic change of the orbital moment upon adding a single electron to the shell, cannot be explained in a single-particle framework. This in itself is strong evidence for many-body effects in our device. The fact that we are able to reproduce this fundamentally non-single particle phenomena using the presented first-principle GW calculations is then an additional strong supporting piece of evidence that these orbital effects have their origin in many-body physics.

Finally, we note that, following similar studies\cite{deshpande2009mott}, we observe the persistence of a non-closing gap at the Dirac field which is not reproduced in our GW calculations. Figure 3(a) shows a representative first-principles calculation for the narrow-gap (9,0) zigzag tube as a function of the magnetic field. The latter is expressed in terms of the magnetic flux ($\phi$) piercing the tube cross section to the flux quantum ($\phi_0$), as we expect a qualitatively similar trend for all narrow-gap tubes, independent from their chirality. The black circles and red squares show respectively density functional theory (DFT, also labeled as 'noninteracting') and GW calculations ('many-body') for the first electron and the first hole ground states. When accounting, from first principles, for the GW self-energy we find a considerable enhancement to the DFT band gap which leads to enhanced orbital magnetic moments and a steeper slope in the ground state energy as a function of magnetic field, essentially restating the effective-mass prediction of Fig. 2(d). Still though, at high enough fields both first-principles and effective-mass GW calculations predict that the electron and hole ground states meet and the transport gap completely closes ($\Delta_{res} = 0$). Figure 3(b) shows the ground state energy of the first electron and the first hole for device A to higher magnetic fields. It can be seen that at 9 T ($\approx 4\times10^{-3}$ $\phi/\phi_0$) the two ground states reach the closest point before diverging at higher fields. Indeed, all four devices show the presence of a non-closing gap at higher fields suggesting an additional contribution to the gap beyond the GW enhancement of the zero-field gap (see Supplemental Materials\cite{SUPP}). We extract residual gaps (at the Dirac field) of $\Delta_{res} = 34, 38$ meV and noninteracting gaps (change in gap energy from B = 0 T to B = 9 T) of $E = 42, 17$ meV for devices A and B, respectively, having diameters of 1.1 and 3 nm.  
Two paradigms have been proposed to explain the presence of this residual gap at the Dirac field in ultra clean carbon nanotube devices, namely, the Mott insulator\cite{kane1997coulomb, nersesyan2003coulomb, odintsov1999universality, lin1998correlation, krotov1997low, balents1997correlation, deshpande2009mott} and the excitonic insulator\cite{rontani2014anomalous, Varsano2017}. Our present study lacks the statistics required to differentiate the two paradigms which predict specific scalings with the nanotube diameter. However, we have shown here that direct measurements of the nanotube diameter are required as interactions in small band gap nanotubes result in enhanced orbital magnetic moments and discrepancies in inferred nanotube diameters.

We have investigated observations of anomalous orbital magnetic moments in ultra-clean carbon nanotube quantum dots. We find that the orbital magnetic moment is up to seven times larger than expected from the semiclassical estimates. We analyze the possible influences on the orbital magnetic moment and find that the simplest corrections do not explain our results.  We instead build a GW corrected effective mass model, supported by first-principle results, and find good agreement with our experimental orbital magnetic moment results. Our measurements suggest that the gapped electronic structure of nominally-metallic CNTs is strongly modified by interaction-driven phenomena. These interactions are rapidly screened by adding a few electrons onto the CNT, which is reflected in the orbital magnetic moment. We note the presence of a non-closing transport gap at higher magnetic fields which falls outside the scope of our developed model but highlights further interaction driven phenomena. Our results emphasize the importance of interactions in ultra-clean carbon nanotube quantum dots and provide the first steps toward closing similar longstanding open questions in low temperature transport studies. 

\textit{Acknowledgments} - The authors thank Herre van der Zant and Shahal Ilani for insightful discussions. We also acknowledge financial support by the Dutch Organization for Fundamental research (NWO/FOM). This work was supported in part by European Union H2020-EINFRA-2015-1 program under grant agreement No. 676598 project “MaX–Materials Design at the Exascale”. E.M. and L.A. acknowledge support from the National Science Foundation under Grant No. 1151369. D.V., E.M. \& M.R. acknowledge PRACE for awarding them access to the Marconi system based in Italy at CINECA (Grant No. Pre14\_3622).

\bibliography{Island_Interactions_ref}
\clearpage
\include{Island_Interactions_supp}

\end{document}

%% file: Island_Interactions_supp.tex
\onecolumngrid

\begin{center}
\textbf{\large Supplemental Material: Interaction-driven giant orbital magnetic moments in carbon nanotubes}
\end{center}
\setcounter{equation}{0}
\setcounter{figure}{0}
\setcounter{table}{0}
\setcounter{page}{1}
\setcounter{section}{0}

\begin{figure}[h]
\centerline{\includegraphics[width=6in]{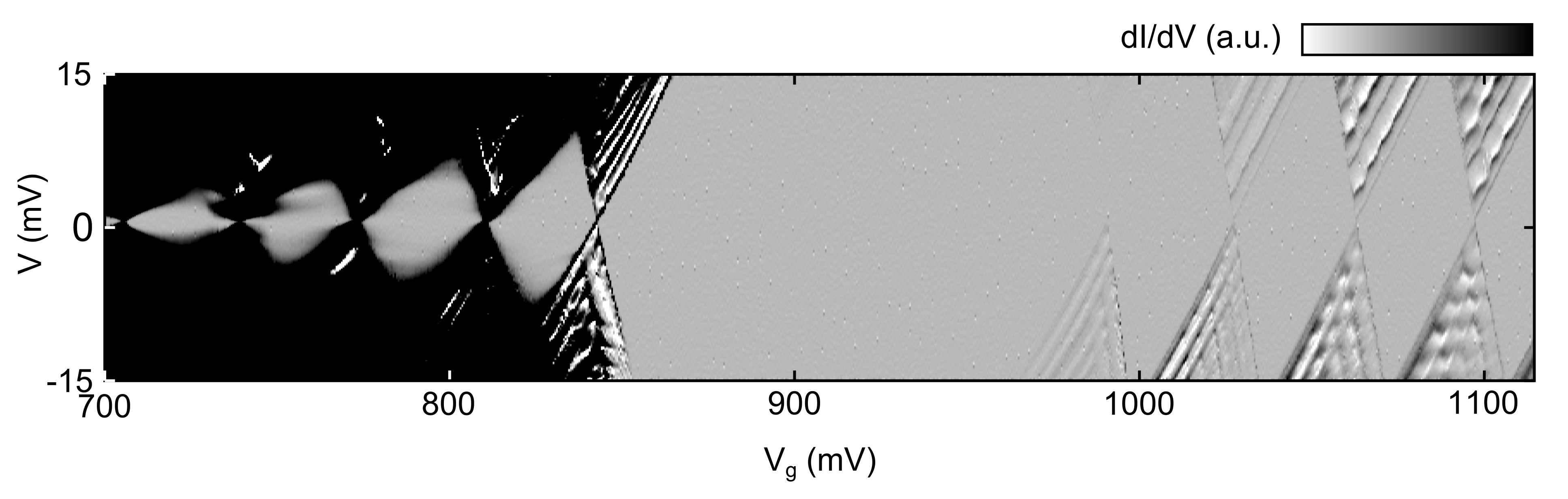}}
\justify \textbf{Figure S1.} \textbf{Low filling stability plot for device A}. Stability plot (calculated d$I$/d$V$ as a function of bias ($V$) and gate ($V_g$) voltage at low electron and hole filling. We estimate the zero field gap by first subtracting the average of the electron (20 meV) and hole (10 meV) addition energies from the $\alpha$-corrected energy between the first electron and hole. For device A we calculate a zero field gap of 76 meV. 
\end{figure}

\twocolumngrid

\section{Modifications to the semiclassical model and magnetic moment of a Wigner molecule}
\subsection{Magnetic field dependent quantum dot length}
It has been shown that the orbital magnetic moment in carbon nanotube quantum dots decreases with each subsequent shell as carriers loose circumferential velocity.\cite{jespersen2011gate} Specifically the orbital magnetic moment scales with the band gap ($E_g$) of the nanotube in the following way:
\begin{equation}\label{EQN1}
\mu_{orb}=\frac{ev_FD}{4\sqrt{1+\left(\frac{2\epsilon_N}{E_g}\right)^2}} 
\end{equation}
with \[ \epsilon_N=\hbar v_FN\pi/L \]
where $\epsilon_N$ is the hard-wall confinement energy, $L$ is the nanotube length, and $N$ is the $N$-th longitudinal mode of the quantum dot. The first few shells will not deviate greatly from the semi-classical value as $\epsilon_N \ll E_g$. However, a further consideration is the change in the size of the quantum dot with increased electron filling which has been speculated as the cause of a large deviation in magnetic moments at low filling\cite{jespersen2011gate}. Typically, the length of the quantum dot is taken to be the physical length of the suspended section of nanotube but this overestimates the size of the quantum dot which is defined by the confinement potential that is sandwiched by the two depletion regions near the source and drain contacts (Supplementary Figure S2). With increased electron filling, the depletion regions become smaller as electrons compete for space on the tube. This change in effective length can be estimated from the gate capacitance extracted from the width of the Coulomb diamonds for each electron, $C_g=e/\Delta V_g$(Ref. \onlinecite{island2011ultra}). The length is then estimated from the wire-plane capacitance model $C/L=2\pi\epsilon /(cosh^{-1}(h/r))$, where $h$ is the distance from the tube to the back-gate, and $r$ is the tube radius. $C_g$ is modeled as two such capacitors in series; one for the oxide of thickness of 185 $nm$ and one for the suspended height (vacuum) of 170 nm. For the first electron we estimate a dot size of 512 nm. This size is increased to 560 nm for the fourth electron. Taking into account this length dependence in Eqn. \ref{EQN1}, the largest deviation from the semi-classical result is $10^{-3}$ meV/T which cannot account for our observations.

\subsection{Magnetic field dependent charging energy}
Assuming the classical estimate for the orbital magnetic moment, we calculate the enhancement to this value by estimating the change in the confining potential with magnetic field. Supplemental Figure S2 shows an energy diagram of device A for the first electron, i.e. a p-doped nanotube between two metal electrodes at finite gate voltage. The zero field gap (76 meV) changes by 0.68 meV/T ($2\mu_{orb}$) or roughly 0.1\%/T with magnetic field. Assuming the charges on the gate do not significantly rearrange in response to the change in gap, the depletion region should scale linearly with the gap. This corresponds to a 0.2 nm/T ($94 nm \times 0.1$\%/T for both sides) change in the depletion regions with magnetic field and therefore a total change in the charging energy of roughly 0.2 nm/T / 512 nm $<$ 0.1 \%/T. The charging energy for the first electron is estimated to be 20 meV (from the diamond height) giving an enhancement of, at most, 0.02 meV/T ($20 meV \times 0.1$\%/T). Together with the magnitude of the classical orbital magnetic moment, this gives a total of 0.30 meV/T. 

\begin{figure}
\centerline{\includegraphics[width=\linewidth]{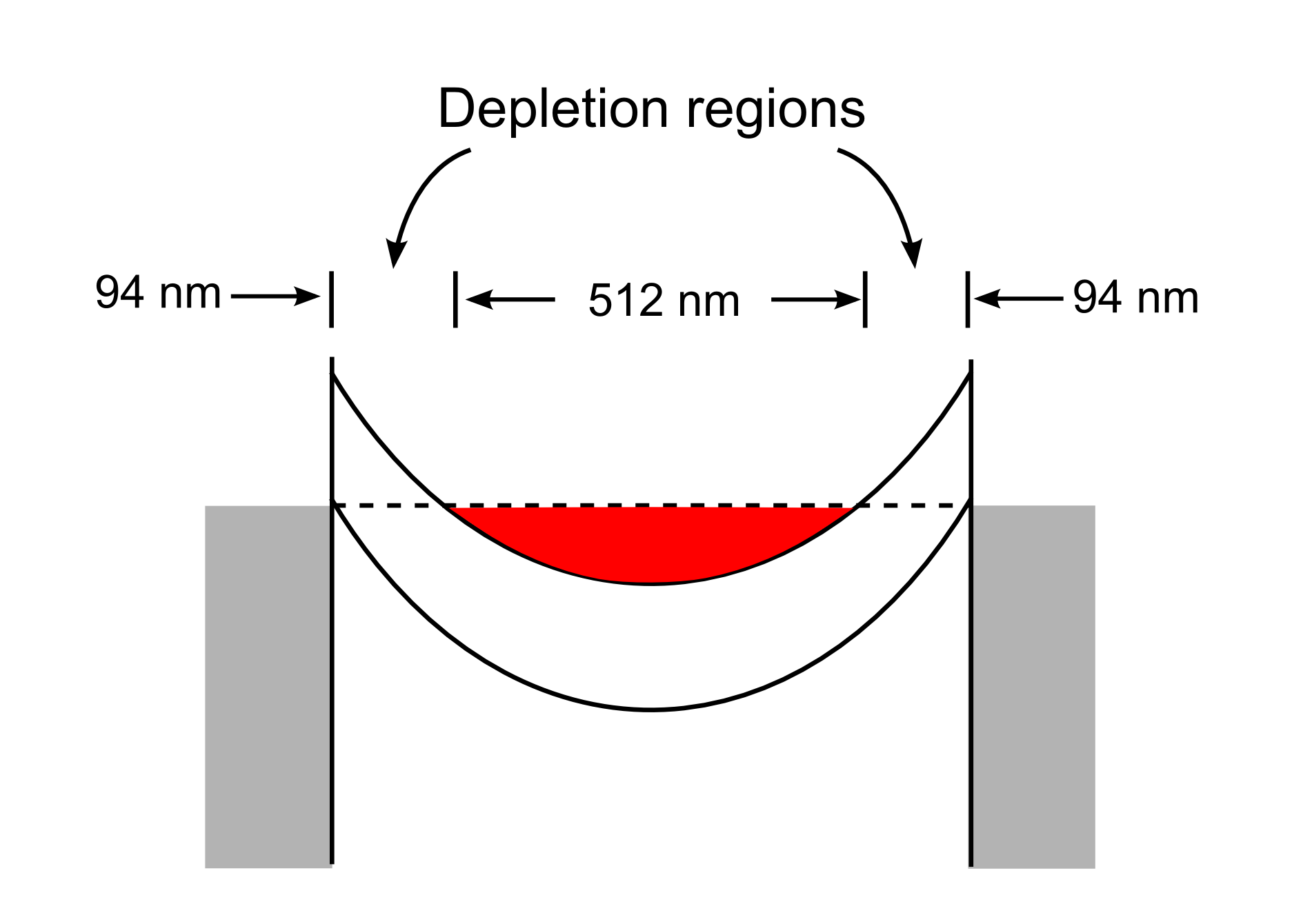}}
\justify \textbf{Figure S2.} \textbf{Estimated change in the charging energy with magnetic field}. Schematic representation of device A showing the depletion regions on either side of the quantum dot filled with electrons (red portion). Assuming the depletion regions scale linearly with the gap, we estimate a change in the charging energy of 0.1 \%/T resulting in, at most, an enhancement to the orbital magnetic moment of 0.02 meV/T.   
\end{figure}

\subsection{Orbital magnetic moment of a Wigner molecule}
We can make an estimate of the orbital magnetic moment for electrons in the Wigner regime by adjusting the semi-classical orbital magnetic moment to account for changes in the Fermi velocity $v_F'=v_Fcos \theta$ with electron filling:
\begin{equation}
\mu_{wig}=\mu_0cos \theta
\end{equation}
with 
\begin{equation}
cos \theta = \frac{k_\parallel}{\sqrt{k_\perp^2+k_\parallel^2}}
\end{equation}
where $k_\perp=E_g/(2\hbar v_F)$ is calculated directly from the estimated band gap, and $k_\parallel=n/L$ where $n$ is the electron number and $L$ is the length estimated from the gate capacitance as above. Supplemental Figure S3(a) shows how $k_\parallel$ increases with electron filling. From this we calculate an orbital magnetic moment of .05 meV for the first electron and an increase to 0.19 meV for the fourth electron. We plot these results along with the experimental data in Supplemental Figure S3(b). It can be seen that the Wigner model also does not suffice to explain the magnitude or trend of the experimentally observed d$E$/d$B$.

\onecolumngrid

\begin{figure}
\centerline{\includegraphics[width=6in]{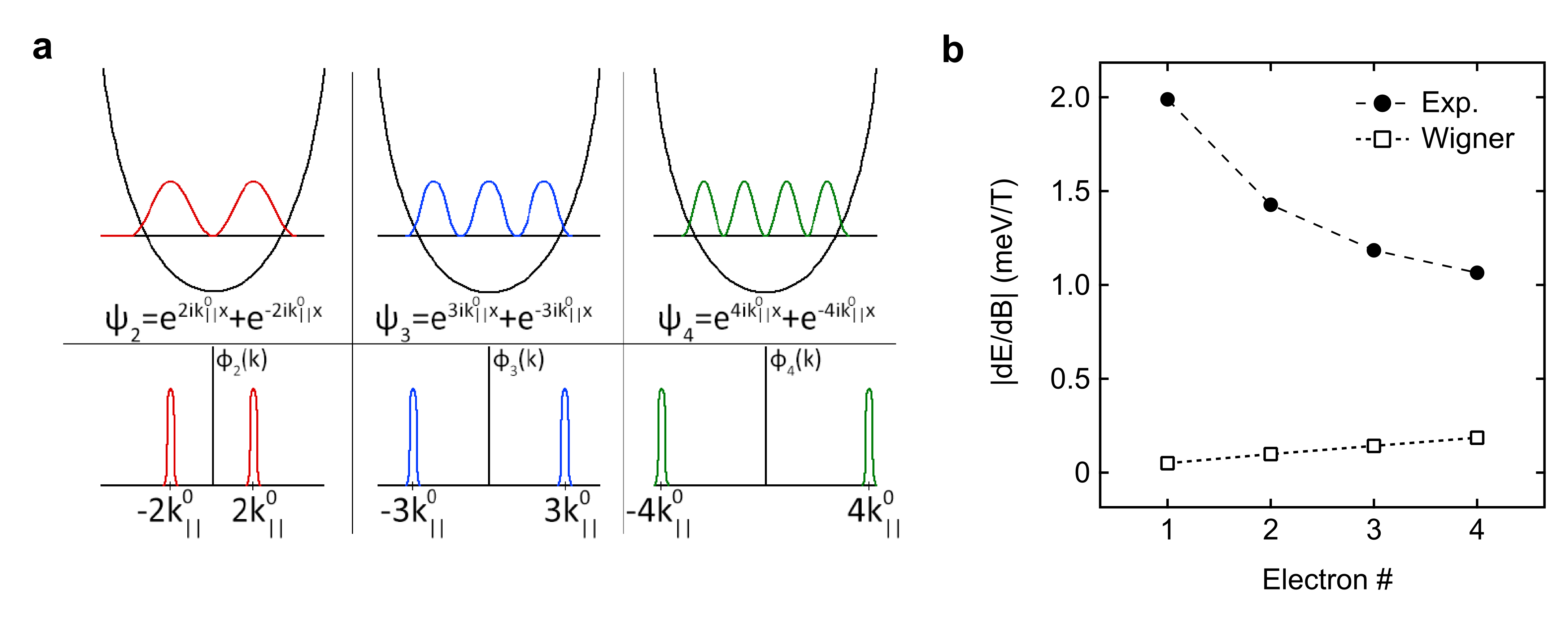}}
\justify \textbf{Figure S3.} \textbf{Orbital magnetic moment of a Wigner molecule.} (a) Schematic of the ground state wavefunctions (2, 3, and 4, electrons) representing 3 subsequent fillings of the nanotube quantum dot (top panels) and corresponding momentum distribution functions as a function of parallel wavevector ($k_{\parallel}$). The increase in the parallel wavevector for subsequent filling leads to larger orbital magnetic moments. (b) The orbital magnetic moment plotted as a function of electron filling. The filled circles show the experimental measured data and the open squares show the estimated orbital magnetic moments from the simple Wigner model. The model fails to reproduce the magnitude or the trend of the results.   
\end{figure}

\twocolumngrid

\section{Measurement summary for devices B, C, and D}

We measured three additional devices (labeled B, C, and D). For device B we have measured directly the diameter of the measured nanotube and compared the measured orbital magnetic moment with the expectations from the semi-classical model. Supplemental Figure S4 shows a summary of these results. As is the case with device A in the main text, we find a disagreement between the measured orbital magnetic moments and those estimated from the noninteracting model. Supplemental Figure S4(f) shows a clear enhancement of the orbital magnetic moments for the first two electrons. 

\onecolumngrid

\begin{figure}
\centerline{\includegraphics[width=6in]{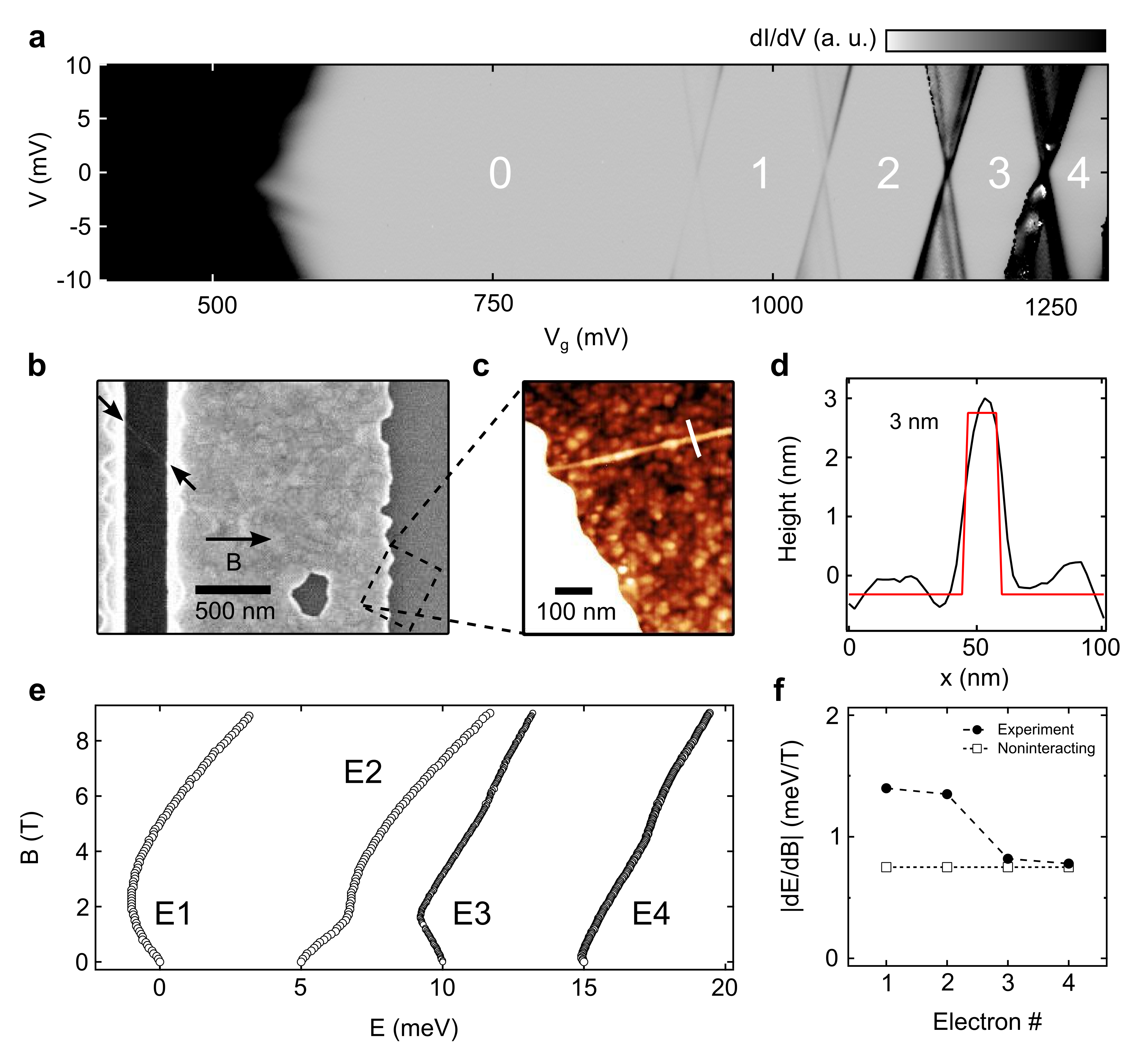}}
\justify \textbf{Figure S4.} \textbf{Summarized experimental results for device B} (a) Stability plot (calculated d$I$/d$V$ as a function of bias ($V$) and gate ($V_g$) voltage at low electron and hole filling. (b) Scanning electron microscopy (SEM) image of the measured nanotube and (c) corresponding atomic force microscopy (AFM) image taken at the location of the box defined by the dashed line. (d) Height profile of the AFM scan taken at the location of the white line in panel (c). (e) The ground state energy for the first four electrons plotted as a function of magnetic field. (f) The magnitude of the orbital magnetic moment for the first four electrons. The filled circles show the experimental results (extracted near 8 T to minimize the influence of the low field Dirac field crossing). The open squares show the estimates from the semi-classical model given the measured nanotube diameter in panel (d). 
\end{figure}

\twocolumngrid

Measurement summaries for devices C and D are shown in Supplemental Figure S5. While we were not able to directly measure the nanotube diameters for these tubes, the trend in the magnitude of the orbital magnetic moment as a function of electron filling is consistent with devices A and B. Note that the fabrication for device D was slightly different and follows the method presented in Ref. \onlinecite{aspitarte2017giant}. The magnitude for both devices decreases monotonically with filling. We note that the orbital magnetic moment for the first electron for device C estimates a diameter greater than the collapse threshold of 5.1 nm for single walled nanotubes. 

\onecolumngrid

\begin{figure}
\centerline{\includegraphics[width=6in]{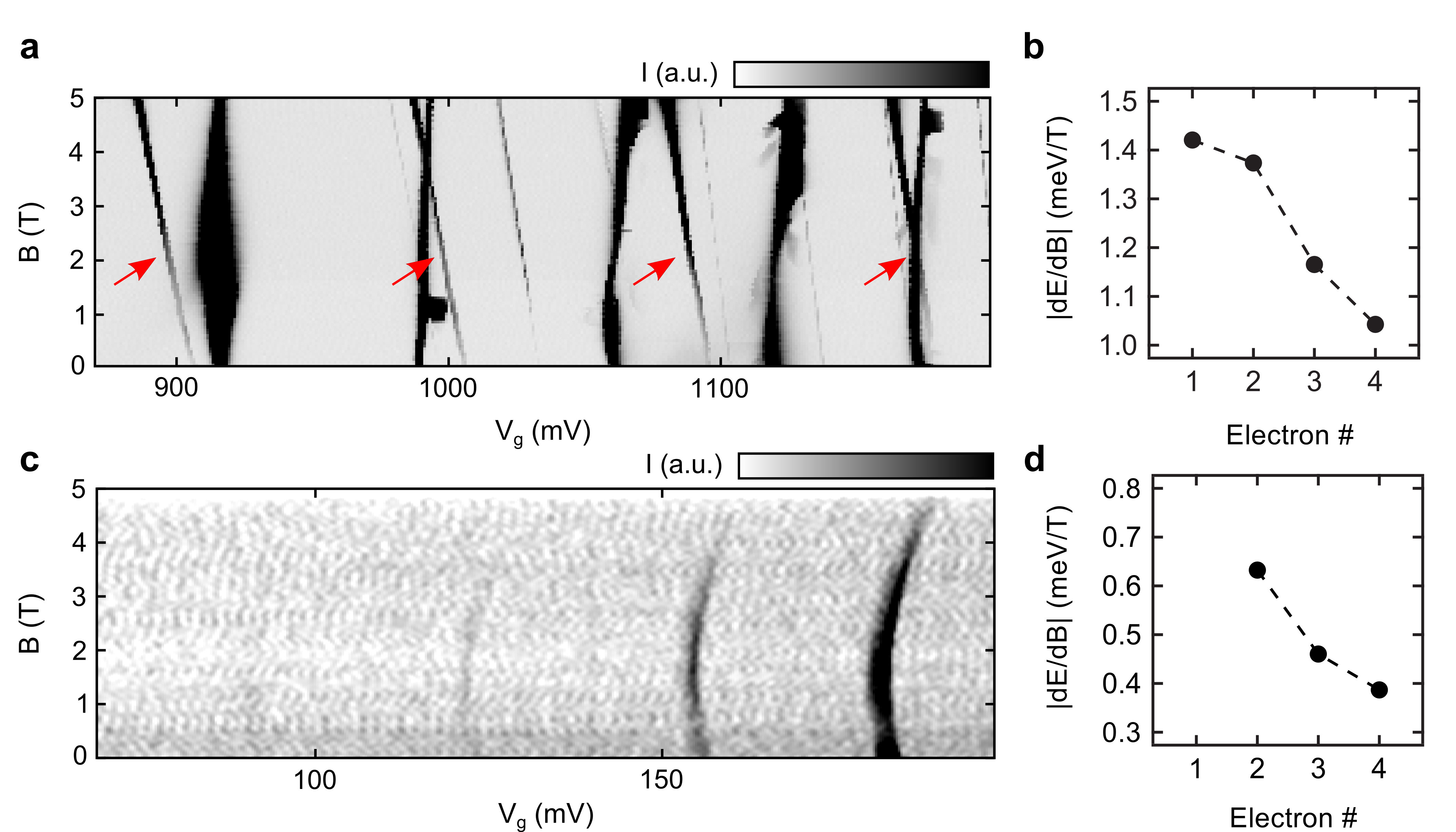}}
\justify \textbf{Figure S5.} \textbf{Measurement summary for devices C and D.} (a) Measured current ($I$) as a function of magnetic field ($B$) and gate voltage ($V_g$) for the first four electrons for device C. The ground states of interest are noted by the red arrows. (b) Magnitude of the orbital magnetic moment for the first four electrons for device C. (c) Measured current ($I$) as a function of magnetic field ($B$) and gate voltage ($V_g$) for the first four electrons for device D. (d) Magnitude of the orbital magnetic moment for electrons 2-4 for device D. The signal to noise ratio was too small to extract the first electron. 
\end{figure}

\twocolumngrid

Supplemental Figure S6 shows the high field behavior for the additional devices (B-D). As in the case of device A, the additional devices show a non-closing gap at the Dirac field where the transport gap should completely close.

\onecolumngrid

\begin{figure}
\justify \centerline{\includegraphics[width=6in]{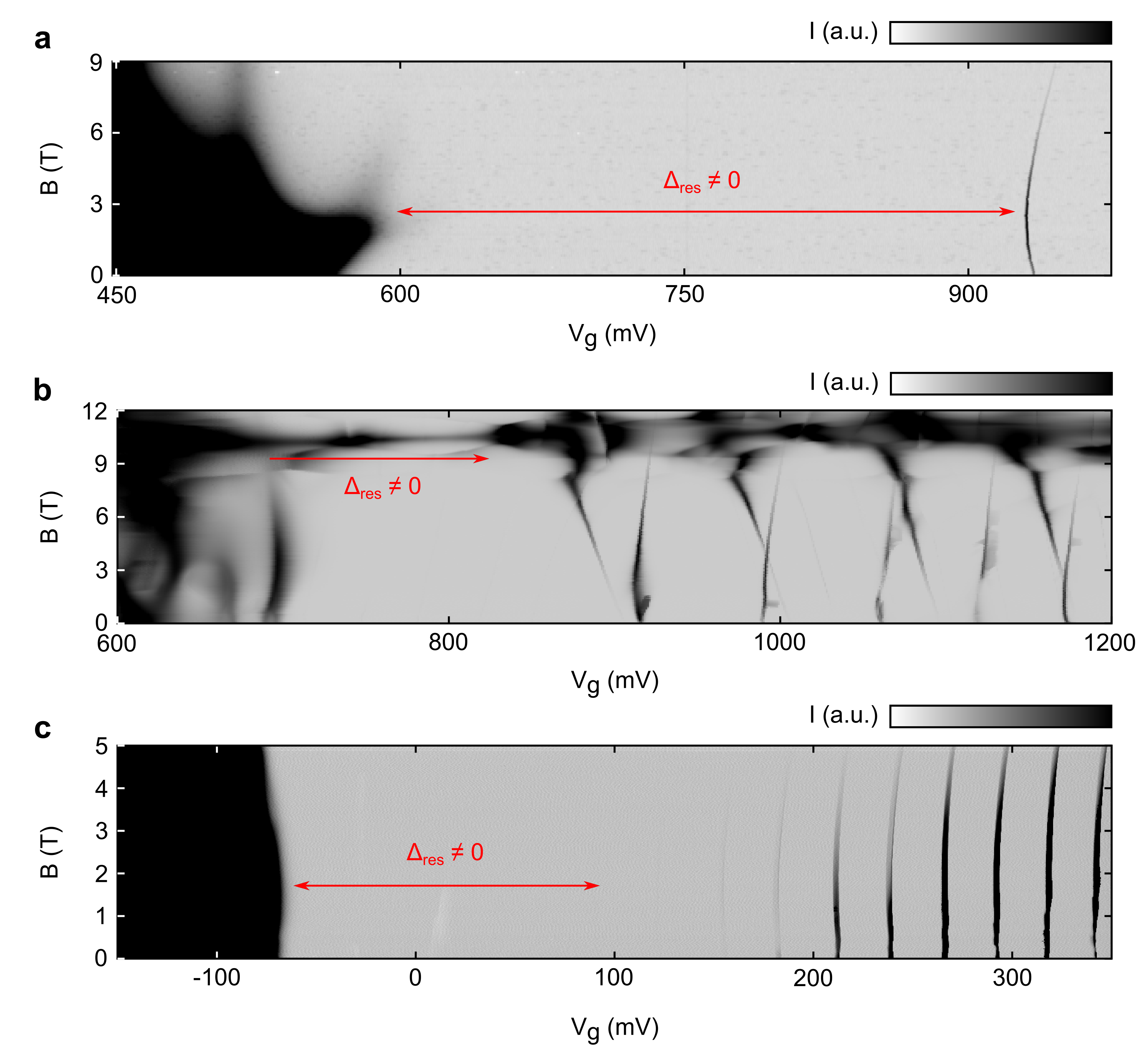}}
\textbf{Figure S6.} \textbf{High magnetic field behavior for devices B, C, and D.} (a) Measured current ($I$) as a function of magnetic field ($B$) and gate voltage ($V_g$) for device B. The red arrow shows the non-closing transport gap. (b) Measured current ($I$) as a function of magnetic field ($B$) and gate voltage ($V_g$) for device C. (c) Measured current ($I$) as a function of magnetic field ($B$) and gate voltage ($V_g$) for device D.
\end{figure}

\twocolumngrid

\section{Computation of the orbital magnetic moment within the $GW$ scheme}

A summary of the central prediction from our calculations and its experimental signature is shown in Figure S7. Figure S7(a) and (b) show the calculated electron transport gap and orbital magnetic moment, respectively, as a function of the chiral angle for a nanotube with a radius of $R=0.55$ nm. This calculation assumes no mechanical twist or axial strain in the nanotube. The effect of many-body interactions is added via many-body perturbation theory which does not describe the possible excitonic insulator phase. Many-body interactions lead to not only a wider transport gap but also an enhanced orbital magnetic moment across all the allowable chiral angles. This very general prediction will manifest itself in magneto-transport measurements (Figure S7(c-d)) as a steeper negative (positive) slope in the first electron (hole) ground state energy as a function of magnetic field as compared with the noninteracting case. The inset diagrams in Fig. S7(c-d) show this schematically for a zigzag (chiral angle $\eta=0$ rad) and a chiral (chiral angle $\eta\approx\pi/8$ rad) nanotube. In either case, the noninteracting (dashed curves) slopes are shallower than the many-body (solid curves) slopes, implying that such nanotubes will exhibit a dramatically enhanced orbital magnetic moment. We note that we did not explore the temperature dependence of the orbital magnetic moement in this work but because the enhancement is given by the many-body self-energy correction, it should follow the temperature dependence of the self-energy which is mainly due to the temperature dependence of the occupied valence band states. Therefore, we would expect a smearing of the many-body enhancement of the magnetic moment at the temperature scale corresponding to the full transport gap (76 meV for device A). However, this would likely require measurements up to quite high temperatures which would make observation of the orbital moment difficult as the temperature broadening would mask the level shifts with magnetic field. 

\onecolumngrid

\begin{figure}
\centerline{\includegraphics [width=6in]{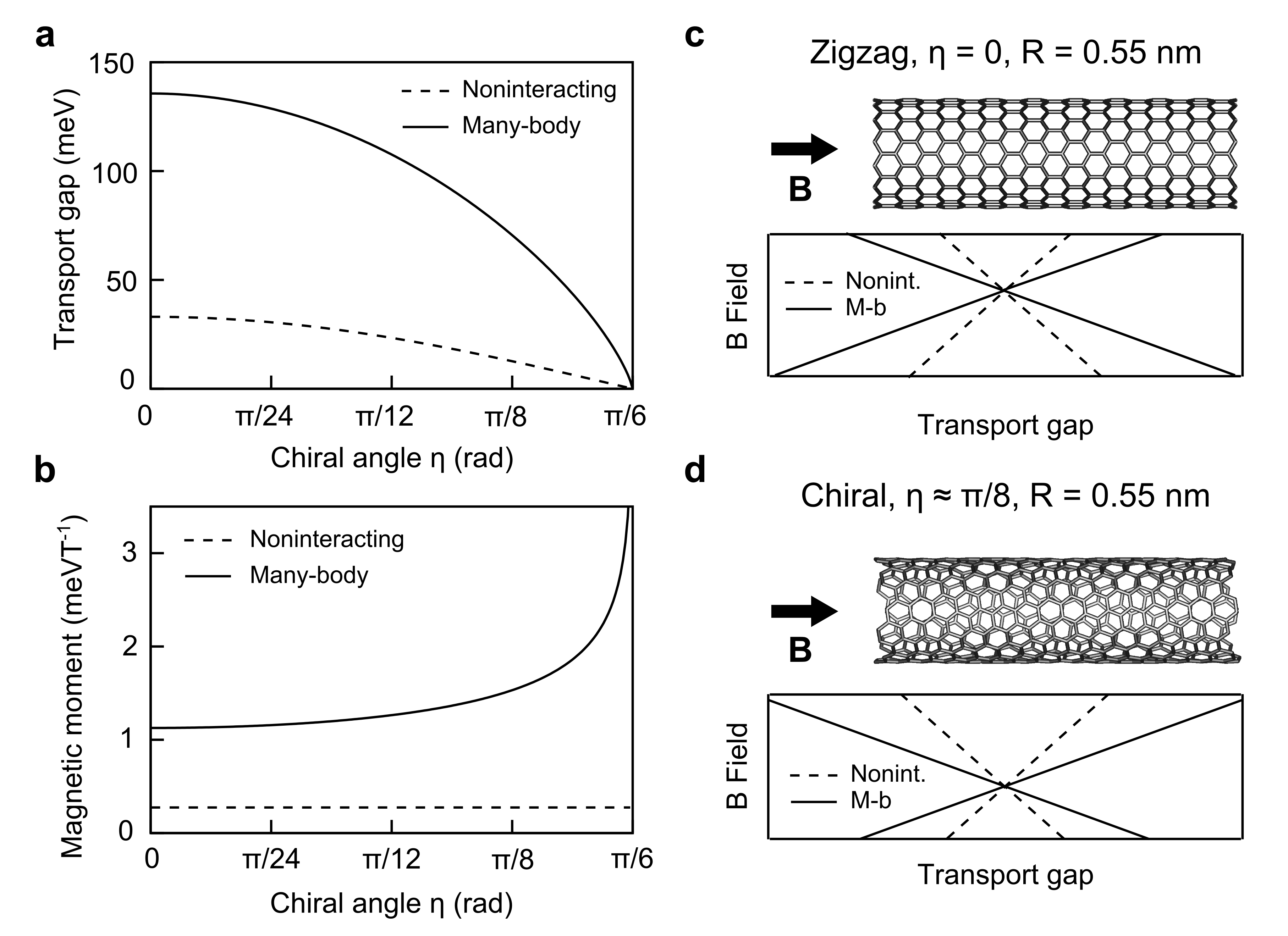}}
\justify \textbf{Figure S7.} \textbf{Predicted giant enhancement of the orbital magnetic moment and accompanying experimental signatures.} (a) Transport gap plotted as a function of the nanotube chiral angle ($\eta$). The solid curve shows the predictions from our effective mass GW calculations ($R = 0.55$ nm). The dashed curve shows the single particle noninteracting results. A sizable enhancement to the transport gap is predicted across the allowable chiral angles. (b) Magnetic moment ($\mu_{orb}$) plotted versus the chiral angle ($\eta$). Our central prediction shows an enhancement to the orbital magnetic moment for all allowable chiral angles. (c) Model showing a zigzag nanotube with a chiral angle of $\eta=0$ and a radius of $R=0.55$ nm. Lower inset shows a diagram of the first hole and electron ground states as a function of magnetic field. The experimental signatures of the predictions in (a) and (b) are a widened transport gap as a result of many-body interactions (solid line) and a steeper slope ($\mu_{orb}=$d$E/$d$B$) of the single electron and hole ground state as a function of axial magnetic field. The dashed line shows the noninteracting case. (d) Model showing a chiral nanotube with $\eta\approx\pi/8$ and a radius of $R=0.55$ nm. Lower panel shows that although the transport gap is smaller, an enhancement to the transport gap and $\mu_{orb}$ is still predicted.
\end{figure}

\twocolumngrid

We compute the orbital magnetic moment of the $N$th electron
added to the carbon nanotube,
$\mu_{\text{orb}}$, as the slope of the corresponding quasiparticle 
energy, $\varepsilon_N$,
versus the magnetic field 
parallel to the tube axis, $B$:
\begin{equation}
\mu_{\text{orb}} = 
- \left(\frac{\partial \,\varepsilon_N}{\partial B}\right)_{B=0}.
\end{equation}
Here $\varepsilon_N$ is 
the difference between the energies
of the interacting many-body
ground states with respectively $N$ and $N-1$ electrons, 
\begin{equation}
\varepsilon_N=E_{\text{GS}}(N)-E_{\text{GS}}(N-1), 
\end{equation}
and we neglect terms in the Hamiltonian that explicitely depend on spin.
The quasiparticle energy differs 
from the noninteracting energy, $\varepsilon_{0N}$, by 
the self-energy, $\Sigma_N$, which takes into account at all orders
the energy shift due to the interaction 
with the other electrons filling in the tube:
\begin{equation}
\varepsilon_N = \varepsilon_{0N} + \Sigma_N.
\end{equation}
The fundamental transport gap is 
\begin{equation}
E_{\text{g}}=2(\varepsilon_0 + \Sigma),
\end{equation}
where we have exploited the electron-hole symmetry, 
placed the origin of the energy axis at the
Dirac point, and dropped the index 1 in $\varepsilon_{01}$.
For the first electron addition, one has
\begin{equation}
\mu_{\text{orb}} = \mu_0 - 
\left(\frac{\partial \Sigma}{\partial B}\right)_{B=0},
\label{eq:mu}
\end{equation}
where   
$\mu_0= - (\partial \varepsilon_0/\partial B)_0$ is 
the noninteracting magnetic moment, which is given
by the semiclassical formula (1) of main text.
This shows that the task of computing $\mu_{\text{orb}}$ amounts to 
evaluate the dependence of $\Sigma$ on $B$. 

Here we derive $\Sigma$ within the screened Hartree-Fock 
approximation\cite{Hedin1969}, also known as $GW$,
by adopting two different methods.
The $GW$ calculation from first principles, which is 
the method of choice as it is
void of free parameters, is especially demanding  
and hence limited to small systems. 
The largest tube we are able to treat  
within the $G0W0$ first-principles scheme\cite{Onida2002}
is the $(9,0)$ zigzag carbon nanotube, which is a member of 
a paradigmatic class of narrow-gap tubes. 
This we use to validate a second
approach, a $\mathbf{k\cdot p}$ screened static Hartree-Fock 
method\cite{Ando1997}, which allows 
to cope with tubes of any size.
In the following we provide details on both methods as well as
on the validation procedure.

\subsection{Details of the first-principles $GW$ calculation}

Ground state calculations for the (9,0) carbon nanotube 
are performed by using a density functional theory (DFT) approach, 
as implemented in the Quantum ESPRESSO package\cite{Giannozzi2009}.
The local density  approximation (LDA) PZ81 parametrization\cite{Perdew1981} 
is adopted together with plane wave basis set and norm-conserving 
pseudopotentials to model the electron-ion interaction. 
The kinetic energy cutoff for the wave functions is set to 90 Ry. 
The Brillouin zone is sampled using a 150 $\times$ 1 $\times$ 1 
$k$-point grid. The supercell size perpendicular to the tube 
is set to 38 Bohr and checked to be large enough to avoid spurious 
interactions with its replica.

The structure of the tube is relaxed allowing all carbon positions 
to change until the forces acting on all atoms become less than 
5.10$^{-3}$ eV$\cdot${\AA}$^{-1}$. The diameter and lattice constant of 
the optimized structure are respectively
7.101 {\AA} and 4.224 {\AA}. The bond lengths between 
neighbouring carbon atoms are $d_1=$ 1.404 {\AA} and $d_2 =$ 1.416 {\AA} 
in good agreement with full-potential linear muffin-tin orbital 
FP-LMTO calculations\cite{Miyake2005}.

Many-body perturbation theory
calculations\cite{Hedin1969,Onida2002} 
are performed using the Yambo code\cite{Marini2009}.
Many-body corrections to Kohn-Sham eigenvalues are calculated within 
the $G0W0$ approximation to the self-energy operator $\Sigma$, the 
dynamic dielectric function being obtained within the plasmon-pole 
approximation\cite{Godby1989}.
The $G0W0$ correction is calculated using a kinetic energy cutoff of 
77.5 Ry for the evaluation of the exchange part of the self energy and 
6 Ry for the screening matrix size. The integration of the self energy 
is accomplished over 700 unoccupied bands. 
In order to speed up the convergence 
with respect to the empty states the technique by 
Bruneval and Gonze\cite{Bruneval2008} is adopted. 
A cutoff in the Coulomb potential, 
in the direction perpendicular to the nanotube axis, is introduced 
in the $G0W0$ calculation to eliminate the spurious interactions 
along the non periodic direction and hence
simulate an isolated nanotube\cite{Rozzi2006}.

The effect of the magnetic field parallel 
to the nanotube axis on the electronic structure of the 
ground state (eigenvalues and eigenfunctions) is investigated following 
the method by Sangalli and Marini\cite{Sangalli2011}.

\subsection{Details of the effective-mass $GW$ calculation}

Within the $\mathbf{k\cdot p}$            
approximation\cite{Ando1997,Varsano2017},
the electronic $\pi$-states of carbon nanotubes
are modelled as pseudo-spinors           
obeying a Dirac-like equation.
They are built starting from the Bloch states of graphene 
and then   
folded into the Brillouin zone of the tube, their transverse wave 
vector being quantized as the graphene sheet is rolled to make a cylinder.
In the original $GW$ theory by Ando\cite{Ando1997}
the self-energy $\Sigma$ was given by a sum
over all bands, which was truncated through a smooth cutoff function. Here
we only consider the lowest conduction and
highest valence bands since the error is small for narrow-gap tubes,
as we check below.
Moreover, we account for the screening action of those free charge carriers 
that are injected into the nanotube by 
Coulomb blockade tunneling spectroscopy.
The resulting expression for $\Sigma$ at the band edge is:
\begin{equation}
\Sigma = \frac{e^2}{A\kappa_{\text{r}}}\sum_q
I_0(R\left|q\right|)\, K_0(R\left|q\right|)
\frac{\left|k_{\perp}\right|}{\epsilon(q)
\left[k_{\perp}^2+q^2\right]^{1/2}},
\label{eq:Sigma}
\end{equation}
where $R$ and $A$ are the nanotube radius and length, 
$q$ and $k_{\perp}$ are the longitudinal and trasverse wave vectors,
$\epsilon(q)$ is the static dielectric function,
$I_0(z)$ and $K_0(z)$ are 
modified Bessel functions of 
the first and second kind,\cite{Abramowitz1972} respectively,
and $\kappa_{\text{r}}$ is a renormalization factor that takes 
into account polarization effects due 
to the electrons not included in the 
$\mathbf{k\cdot p}$ description
plus the contribution of the dielectric background.

The transverse wave vector $k_{\perp}$ appearing in (\ref{eq:Sigma})
controls the noninteracting gap, 
$2\hbar v^{\text{graph}}_{\text{F}}\left|k_{\perp}\right|$,
with the noninteracting bands being 
\begin{equation}
\varepsilon_0(k)=\pm \hbar 
v^{\text{graph}}_{\text{F}}\left[k_{\perp}^2 + k^2\right]^{1/2},
\end{equation}
where $v^{\text{graph}}_{\text{F}}$ is graphene's Fermi velocity, 
and plus and minus signs
refer to conduction and valence bands, respectively.
In the present case $k_{\perp}$ is small and given by the sum of two terms,
\begin{equation}
k_{\perp} = \tau \frac{{\cal C}}{\hbar v^{\text{graph}}_{\text{F}} R^2}
+ \frac{e\pi R}{ch}B,
\label{eq:k_perp}
\end{equation}
the first one depending on tube chirality and curvature, 
the second one on the magnetic flux piercing the cross section.
Here $\tau$ is the valley index, with $\tau=1$ or -1 for valleys K and K$'$,
respectively, and ${\cal C}$ is a numerical factor that depends on the
chirality angle $\eta$ as\cite{laird2015quantum}
\begin{equation}
{\cal C}={\cal C}_0\cos{3\eta},
\end{equation}
where ${\cal C}_0=$ 0.5 eV$\cdot${\AA}$^2$ and $\eta$ varies
between $\eta=0$ (zigzag tube) and $\eta=\pi/6$ 
(armchair).  
The magnetic flux displaces $k_{\perp}$ in reciprocal space
through the Aharonov-Bohm effect, which has an opposite effect
in the two valleys: 
the gap increases with $B$ in one valley 
whereas it decreases in the other one.

The dielectric function accounts for both inter and
intraband contribution to the polarization, respectively $\Pi_{\text{inter}}$
and $\Pi_{\text{intra}}$,
\begin{equation}
\epsilon(q) = 1 + \frac{2e^2}{\kappa_{\text{r}}}
I_0(R\left|q\right|)\, K_0(R\left|q\right|)
\Big[ \Pi_{\text{intra}}(q) + \Pi_{\text{inter}}(q)\Big].
\end{equation}
Screening is poor in the undoped nanotube\cite{Ando1997},
since $\Pi_{\text{inter}}(q)$ vanishes as $q\rightarrow 0$ 
and there is no intraband
contribution, $\Pi_{\text{intra}}=0$.
Here we take the form
$\Pi_{\text{inter}}(q)=A_{\text{ansatz}}(Rq)^2$
with $A_{\text{ansatz}}=50\cdot (\pi\hbar v_{\text{F}}^{\text{graph}})^{-1}$,
following a previous study of a small tube that has been
validated from first principles\cite{Varsano2017}.
On the contrary,
screening is very effective in the doped tube,
as the intraband contribution to the polarization, $\Pi_{\text{intra}}$,  
overwhelms the interband term, $\Pi_{\text{inter}}$.
In fact,
the intraband polarization is the Lindhard function of 
a one-dimensional metal\cite{Giuliani2005} with double valley degeneracy,
\begin{equation}
\Pi_{\text{intra}}(q) = \frac{ 2 }{ \pi\hbar 
v_{\text{F}}^{\text{graph}} } \sum_{\tau} 
\frac{  \left| k_{\perp}(\tau) \right| }{ 
\left|q\right| } \log{ \left| \frac{2k_{\text{F}} + q}
{2k_{\text{F}}-q} \right| }.
\end{equation}
Here we parametrize the charge injected into the
conduction band through the Fermi wave vector, $k_{\text{F}}$ 
(note that this is unrelated to $v^{\text{graph}}_{\text{F}}$). In order to simulate device A, we take as the Fermi wave vector of the Nth electron added to the conduction band, $k_F =  0.5 \times 10^{-5}(N - 1) (2 \pi) / a$, with $a = 0.246$ nm being the lattice constant for graphene.
Since $\Pi_{\text{intra}}$ is singular for $q\rightarrow 2k_{\text{F}}$
and proportional to the density of states for $q\rightarrow 0$,
$\epsilon(q)$ diverges in both limits ($K_0$ diverges
for $q\rightarrow 0$).

In principle, the expression (\ref{eq:Sigma}) 
for $\Sigma$ only applies to the
first electron added to the conduction band edge at $k=0$. 
Since---in a noninteracting picture---relevant electron additions 
occur only within the first shell, we neglect the
dependence of $\Sigma$ on $k$ and include  
the effects of charging into $\Pi_{\text{intra}}$. Moreover, 
we neglect the effect of Coulomb blockade.

To proceed, we rewrite (\ref{eq:Sigma}) in
the thermodynamic limit, $A\rightarrow \infty$,
\begin{equation}
\Sigma = \frac{e^2\left|k_{\perp}\right|}{\pi\kappa_{\text{r}}}I,
\label{eq:Sigma_bis}
\end{equation}
with
\begin{equation}
I = 
\int_0^{\infty}dz \frac{ I_0(z)K_0(z) }{\tilde{\epsilon}(z)
\left[R^2k_{\perp}^2+z^2\right]^{1/2}},
\label{eq:Sigma_ter}
\end{equation}
and $\tilde{\epsilon}(z) = \epsilon(z/R)$. The integral 
$I$ converges since the
the kernel  
has a logarithmic singularity for $z\rightarrow 0$ (and $k_{\text{F}}=0$) 
and vanishes faster than $1/z$
for $z\rightarrow \infty$.
However, $I$ diverges for vanishing noninteracting gap, $k_{\perp}
\rightarrow 0$.
Since $\Sigma$ only depends on $B$ through $k_{\perp}$ 
[cf.~(\ref{eq:k_perp})],
we may neglect the weak dependence of $I$ on $B$ 
when evaluating $\partial \Sigma/\partial B$,
which provides the important result:
\begin{equation} 
\left|\mu_{\text{orb}}\right| \approx 
\left|\mu_0\right|\left(1 + \frac{\Sigma}{\varepsilon_0}\right)=
\left|\mu_0\right|\left(1 + \frac{e^2}{\pi \kappa_{\text{r}}\hbar
v_{\text{F}}^{\text{graph}}}I\right). 
\label{eq:mu_orb}
\end{equation}

\subsection{Validation from first principles}

To validate the effective-mass approach, we compare the
$GW$ values of $E_{\text{g}}$ obtained respectively from first-principles 
and effective-mass theory for the (9,0) zizgag tube. 
The free parameters of the effective-mass method 
are chosen as follows.
The radius is obtained from DFT structural optimization, giving $R=3.55$ {\AA}, 
and $\eta=0$. Besides, we take the same parameters
$v_{\text{F}}^{\text{graph}}=10^6$ m$\cdot$s$^{-1}$ and
$\kappa_r=2.5$ used to obtain 
$\mu_{\text{orb}}$ for device A 
(the chiral angle $\eta=0.151 \pi$ of device A was 
inferred by the observed Dirac
value of $B$). 
These parameters provide an estimate of
$E_{\text{g}}=$ 200 meV, which reasonably compares with the first-principles
prediction of $E_{\text{g}}=$ 260 meV, since  
$\kappa_{\text{r}}$ includes
environmental screening effects that are absent in the 
first-principles calculation.

Furthermore, we check the effective-mass theory against the
first-principles $GW$ predictions by Spataru\cite{Spataru2012} 
for the zigazag tubes
(10,0) and (17,0). Since these tubes have 
large gaps of the
order of 1 eV, 
terms like $\pm 1/3R$ must be added to $k_{\perp}$ in Eq.~(\ref{eq:k_perp}).
According to Spataru's calculation, the tube (10,0) 
has $R=3.9$ {\AA} and $E_{\text{g}}=1.72$ eV, whereas the tube
(17,0) has $R=6.6$ {\AA} and $E_{\text{g}}= 1.29$. We are
able to match these values by using respectively $\kappa_{\text{r}}=2.2$ 
and $\kappa_{\text{r}}=1.1$ in the effective-mass calculation.
This appears to be a reasonable trend since we expect that contributions 
to the dielectric function due to $\sigma$-$\pi$ band hybridization,
which are mimicked by $\kappa_{\text{r}}$, vanish
for large $R$. In addition, we have checked that the error on $E_{\text{g}}$
due to neglecting higher-energy bands in the effective-mass calculation
is of the order of 10\% for the (10,0) tube and of 7\% for the (17,0), which
shows that the magnitude of the error scales with the size of
the gap. This validates 
our two-band model 
for the present case of narrow-gap tubes.